\documentstyle[11pt,aaspp4]{article}
\font\smfont=cmr7
\font\smtfont=cmr5
\font\smttfont=cmr5

\newfam\smfam

\textfont\smfam=\smfont
\scriptfont\smfam=\smtfont
\scriptscriptfont\smfam=\smttfont
\def\spose#1{\hbox to 0pt{#1\hss}}

\def\kms{\ifmmode {\rm\,km\,s^{-1}}\else
    ${\rm\,km\,s^{-1}}$\fi}
\def\kmsMpc{\ifmmode {\rm\,km\,s^{-1}\,Mpc^{-1}}\else
    ${\rm\,km\,s^{-1}\,Mpc^{-1}}$\fi}

\def\msun{\ifmmode {\rm\,M_\odot}\else ${\rm\,M_\odot}$\fi}
\def\Msun{\ifmmode {\rm\,M_\odot}\else ${\rm\,M_\odot}$\fi}
\def\lsun{\ifmmode {\rm\,L_\odot}\else ${\rm\,L_\odot}$\fi}
\def\Lsun{\ifmmode {\rm\,L_\odot}\else ${\rm\,L_\odot}$\fi}
\def\rsun{\ifmmode {\rm\,R_\odot}\else ${\rm\,R_\odot}$\fi}
\def\Rsun{\ifmmode {\rm\,R_\odot}\else ${\rm\,R_\odot}$\fi}
\def\micron{\ifmmode {\,\mu{\rm m}}\else
    ${\,\mu{\rm m}}$\fi}

\def\cm{{\rm\,cm}}
\def\cm3{\ifmmode {\rm\,cm^{-3}}\else ${\rm\,cm^{-3}}$\fi}

\def\Lya{\ifmmode {\rm\,Ly\alpha}\else ${\rm\,Ly\alpha}$\fi}
\def\Ha{\ifmmode {\rm\,H\alpha}\else ${\rm\,H\alpha}$\fi}
\def\Hb{\ifmmode {\rm\,H\beta}\else ${\rm\,H\beta}$\fi}
\def\Hg{\ifmmode {\rm\,H\gamma}\else ${\rm\,H\gamma}$\fi}
\def\FeII{\ion{Fe}{2}}

\def\OIII{[\ion{O}{3}]}
\def\OII{[\ion{O}{2}]}

\def\ergps{\ifmmode {\rm\,erg\,s^{-1}}\else ${\rm\,erg\,s^{-1}}$\fi}
\def\ergpscm2{\ifmmode {\rm\,erg\,s^{-1}\,cm^{-2}}\else
    ${\rm\,erg\,s^{-1}\,cm^{-2}}$\fi}

\def\eg{{e.g.}}
\def\deg{\ifmmode {^{\circ}}\else {$^\circ$}\fi}
\def\degr{\ifmmode {^{\circ}}\else {$^\circ$}\fi}
\def\degs{\ifmmode {^{\circ}}\else {$^\circ$}\fi}

\def\etal{{et al.~}}

\def\h3Mpc{h^{-3}{\rm Mpc}^3}
\def\Ho{\ifmmode {\rm\,H_0}\else ${\rm\,H_0}$\fi}
\def\hnot{\ifmmode {\rm\,H_0}\else ${\rm\,H_0}$\fi}
\def\h0{\ifmmode {\rm\,H_0}\else ${\rm\,H_0}$\fi}
\def\hnotunit{\ifmmode {\rm\,km\,s^{-1}\,Mpc^{-1}}\else
    ${\rm\,km\,s^{-1}\,Mpc^{-1}}$\fi}
\def\qnot{\ifmmode {\rm\,q_0}\else ${\rm q_0}$\fi}
\def\q0{\ifmmode {\rm\,q_0}\else ${\rm q_0}$\fi}

\def\mic{\ifmmode {\rm\,\mu m}\else ${\rm \mu m}$\fi}

\def\arcsec{\ifmmode {^{\prime\prime}}\else $^{\prime\prime}$\fi}
\def\asec{\ifmmode {^{\prime\prime}}\else $^{\prime\prime}$\fi}
\def\arcmin{\ifmmode {^{\prime}}\else $^{\prime}$\fi}
\def\amin{\ifmmode {^{\prime}}\else $^{\prime}$\fi}

\def\secper{\ifmmode \rlap.{^{s}}\else $\rlap{.}{^{s}} $\fi}
\def\minper{\ifmmode \rlap.{^{m}}\else $\rlap{.}{^m} $\fi}
\def\magper{\ifmmode \rlap.{^{m}}\else $\rlap{.}{^m} $\fi}
\def\farcs{\ifmmode \rlap.{^{\prime\prime}}\else
    $\rlap.{^{\prime\prime}}$\fi}
\def\arcsper{\ifmmode \rlap.{^{\prime\prime}}\else
    $\rlap.{^{\prime\prime}}$\fi}
\def\arcmper{\ifmmode \rlap.{^{\prime}}\else
    $\rlap.{^{\prime}}$\fi}
\def\spose#1{\hbox to 0pt{#1\hss}}
\def\simlt{\mathrel{\spose{\lower 3pt\hbox{$\mathchar"218$}}
     \raise 2.0pt\hbox{$\mathchar"13C$}}}
\def\simgt{\mathrel{\spose{\lower 3pt\hbox{$\mathchar"218$}}
     \raise 2.0pt\hbox{$\mathchar"13E$}}}

\def\refindent{\par\noindent\parskip=2pt\hangindent=3pc\hangafter=1 }

\def\araa{{ARA\&A}}

\def\aj{{AJ}}
\def\apj{{ApJ}}

\def\apjl{{ApJ}}
\def\apjsupp{{ApJS}}
\def\apjs{{ApJS}}

\def\mnras{{MNRAS}}

\def\apjref#1;#2;#3;#4 {\par\pp#1, {#2}, #3, #4 \par}

\begin{document}

\title{Red Quasars and Quasar Evolution: the Case of 
BAL QSO \\ FIRST J155633.8+351758\altaffilmark{1}}
\author{Joan Najita, Arjun Dey, \& Michael Brotherton }
\affil{KPNO/NOAO\altaffilmark{2}, 950 N. Cherry Ave., P.\ O.\ Box 26732, Tucson,
AZ 85726}
\affil{najita, dey, mbrother@noao.edu}

\altaffiltext{1}{Based on observations at the Kitt Peak National Observatory.}
\altaffiltext{2}{The National Optical Astronomy Observatories are
operated by the Association of Universities for Research in Astronomy under
Cooperative Agreement with the National Science Foundation.}

\begin{abstract}

We present the first near-infrared spectroscopic observations of the
radio-loud broad absorption line QSO FIRST J155633.8+351758. The spectrum
is similar to that of a reddened QSO and shows strong emission lines
of \Ha\ and \Hb\ as well as strong \FeII\ emission blends 
near \Hb.  The redshift of the object, measured from the \Ha\ and \Hb\ 
lines, is $z_{\rm BLR}=1.5008\pm0.0007$, 
slightly larger than the redshift of $z_{\rm metal}=1.48$ estimated from 
the broad metal absorption features.  Thus, the broad metal 
absorption features are blue shifted with respect to the systemic 
velocity.  
The width of the H$\alpha$ emission line (${\rm FWHM \approx
4100~km~s^{-1}}$) is typical of that observed in QSO broad-line
regions, but the Balmer decrement (H$\alpha$/H$\beta$$\approx 5.8$) is larger than 
that of most optically selected QSOs.  Both the Balmer decrement 
and the slope of the rest-frame UV-optical continuum independently 
suggest a modest amount of
extinction along the line of sight to the broad-line region 
($E_{B-V}\approx 0.5$ for SMC-type screen extinction at the redshift 
of the QSO). 
The implied gas column density along the line of sight is much less 
than is implied by the weak X-ray flux of the object, suggesting that 
either the broad emission and absorption line regions have a low 
dust-to-gas ratio, or that the rest-frame optical light encounters 
significantly lower mean column density lines of sight than the X-ray emission.  
From the rest-frame UV-optical spectrum, we are able to constrain the 
stellar mass content of the system ($< 3\times 10^{11}\Msun$).  
Comparing this mass limit with the black hole mass estimated 
from the bolometric luminosity of the QSO, we find it possible 
that the ratio of the black hole to stellar mass is comparable to 
the Magorrian value, which would imply that the Magorrian relation is 
already in place at $z=1.5$.
However, multiple factors favor a much larger black hole to stellar mass 
ratio.  This would imply 
that if the Magorrian relation characterizes the late history of QSOs, 
and the situation observed for F1556+3517 is typical of the early 
evolutionary history of QSOs, central black hole masses develop more
rapidly than bulge masses.

\end{abstract}

\section{Introduction}

Some 10-20\% of quasistellar objects (QSOs) show blue shifted, broad absorption 
in high-ionization lines (e.g., \ion{C}{4} and \ion{N}{5}) in their 
rest-frame UV spectra (Weymann et al.\ 1981; Becker et al.\ 2000). 
These broad absorption line QSOs (BALQSOs) may provide 
valuable clues to the origin of characteristic features of QSOs
(e.g., outflow phenomena, obscuring tori) and to the life history of QSOs.  
For example, in BALQSOs, outflow phenomena are readily apparent, 
which makes it easier to study their ejection dynamics for comparison 
with the predictions of dynamical theories (e.g., Murray \& Chiang 1995; 
Scoville \& Norman 1995).
The underlying physics inferred from such comparisons may apply to a 
broader class of QSOs since several lines of evidence suggest that BALQSOs are 
simply normal QSOs viewed along a particular line of sight, i.e., closer to 
edge-on, along a line of sight that skirts an obscuring torus or wind 
from an accretion disk. 
The torus and wind may be related dynamically, e.g., if 
the BAL absorption arises in material ablated off  
the obscuring torus (e.g., Weymann et al.\ 1991; Voit, Weymann \& Korista 1993). 
The evidence in favor of this orientation-dependent unification scenario 
includes the fact that the emission-line properties of BALQSOs are quite similar 
to those of non-BALQSOs (Weymann et al.\ 1991), and  
the higher polarization 
(e.g., Ogle 1997; Ogle et al.\ 1999; Schmidt \& Hines 1999) 
and larger reddening (e.g., Sprayberry \& Foltz 1992; Turnshek et al.\ 1994) 
of BALQSOs compared to non-BALQSOs. 
In particular, spectropolarimetric studies of BALQSOs find 
evidence for at least two lines of sight to the QSO: 
a direct, edge-on line of sight that is perhaps attenuated,
plus a less extincted, scattered (and hence polarized) line of
sight that passes above the disk wind or torus 
(Hines et al. 1995; Cohen et al. 1995; Goodrich \& Miller 1995).
If the incidence rate of BALQSOs among QSOs is indeed determined by orientation, 
then what we learn about the mechanics of ejection based on BALQSOs may 
be applicable to QSO outflows in general.

Following a different line of argument, BALQSOs may represent a particular 
evolutionary phase in the QSO lifecycle and, therefore, 
the fraction of QSOs that are BALQSOs may reflect the 
fraction of time that any given QSO spends in a BAL phase. 
So, for example, the red color of BALQSOs may suggest a large 
line-of-sight extinction and the possibility that we are viewing this object
in an `embryonic' state 
(e.g., Sanders et al.\ 1988; Voit et al.\ 1993). 
Consequently, what we learn from the study of BALQSOs may provide clues 
to the early stages of the life history of QSOs.
As a more specfic example,
in their study of red, high-z QSOs, the large Balmer decrements 
of Hawaii~167 ($z=2.36$; H$\alpha$/H$\beta$=13) and 
Q0059-2735 ($z=1.59$; H$\alpha$/H$\beta$=7.6) led 
Egami et al.\ (1996) to hypothesize that these are young QSOs 
forming in a shroud of massive stars and dust. 

The object of the present study, 
FIRST J155633.8+351758 (hereinafter F1556+3517), 
bears some resemblance to these red QSOs and is itself an extreme 
and unusual member of the BALQSO class.  As summarized by 
Becker et al.\ (1997), it is a rare, iron low-ionization 
BALQSO discovered through optical follow up of the FIRST radio survey,  
and has been proposed as the first radio-loud BALQSO known 
(i.e., the ratio of rest-frame 5 Ghz to 2500\AA\ flux density is $>1000$; 
Becker et al.\ 1997).  
F1556+3517 is also one of only a handful of QSOs to display 
absorption from metastable states of Fe II
(Becker et al. 1997).  Moreover,  
it is very red ($B-K\simeq 6.6$; Hall et al. 1997), redder than almost 
all QSOs at $z\simeq 1-2$, suggesting a large column density of dust along 
our line of sight. 

Like other BALQSOs, F1556+3517 is highly polarized in the rest-frame UV continuum 
($\approx$13\% at 2000\AA; Brotherton et al.\ 1997; cf. Hines \& Schmidt 1997; 
Ogle et al.\ 1999). 
However, unlike other BALQSOs, 
the emission lines and continuua are similarly polarized, and 
the BAL troughs are unpolarized (Brotherton et al.\ 1997). 
As described by Brotherton et al.\ (1997),  
the polarized component is apparently light from the QSO nucleus 
that is scattered into our line of sight by a scatterer located, 
unconventionally, 
outside the broad emission and absorption line regions, producing a similar level of 
polarization in the emission lines and the continuum. 
The unpolarized light in the BAL troughs may be either light from a 
starburst or reddened, unabsorbed QSO light viewed directly 
(i.e., not absorbed in the BAL region).  
In this paper, we use near infrared (rest-frame optical) spectroscopy of 
F1556+3517 
to constrain the extinction toward the BAL region and 
to explore the stellar content of the system.

\section{Observations}

We observed F1556+3517 on U.T.~1997 May 22 and U.T.~1997 June 23 using the
near-IR Cryogenic Spectrometer (CRSP; Joyce et al.\ 1994, 
Joyce 1995) at the Cassegrain
focus of the 4-m Mayall Telescope of the Kitt Peak National
Observatory. During the May run, the spectrograph was configured with
the 300 l/mm grating (\#1) which was used in 2nd order with the $H$
filter to cover the wavelength range $\lambda\lambda 1.526-1.704\mu$m
(dispersion $\approx 7{\rm \AA~pixel^{-1}}$) and in 3rd order with the
$J$ filter for the range $\lambda\lambda 1.1705-1.2785\micron$
(dispersion $\approx 4.2{\rm \AA~pixel^{-1}}$).  Observations of the
nearby G5V star HR6538, made at approximately the same airmass and
immediately following the observations of F1556+3517, were used to
correct the spectrum for telluric absorption and for relative flux
calibration. Our May observations were obtained through a
1\arcsec\ (2.7 pixels) slit, resulting in resolution FWHMs of $\approx
19$\AA\ and 11\AA\ in the $H$ and $J$ bands respectively.  During the
June run, we used the 200 l/mm grating (\#4) which was used in 3rd
order with the $I$ filter to cover
$\lambda\lambda$0.894--1.103\micron\  (dispersion $\approx 8.2{\rm
\AA~pixel^{-1}}$),  in 2nd order with the $J$ filter to cover
$\lambda\lambda$1.067--1.3916\micron\ (dispersion $\approx 12.7{\rm
\AA~pixel^{-1}}$), and in 2nd order with the $H$ filter to cover
$\lambda\lambda$1.447--1.755\micron\ (dispersion $\approx 12.1{\rm
\AA~pixel^{-1}}$). The G1V star HR6064 was observed at roughly the same
airmass as a telluric standard. The June observations were obtained
through a 1\arcsper5 (4 pixels) slit, resulting in resolution FWHMs in
the $I$, $J$ and $H$ bands of $\approx 33$\AA, 51\AA\ and
48\AA\ respectively.  Conditions were photometric on both runs, but
poor seeing and image motion during the June observations resulted in
larger slit losses.  

The data were dark subtracted, flat-fielded, sky
subtracted and coadded using a bad-pixel mask before the spectra were
extracted.  Relative flux calibration for the May and June observations
was performed using HD203856 and HD136754 respectively (Elias \etal
1982). The extracted spectra from the two runs were coaveraged and
scaled to match the broad band photometry reported by Hall \etal
(1997) using the ONIS filter transmission curves. 
The relative flux calibration of the two data sets are in
excellent agreement, and the object shows no evidence for variability
in the broad emission lines between May and June 1997.  The data were
reduced and analyzed using {\it IRAF}\footnote{IRAF is distributed by 
the National Optical Astronomy Observatories,
which are operated by the Association of Universities for Research
in Astronomy, Inc., under cooperative agreement with the National
Science Foundation.}
and custom {\it IDL} routines.

\section{Results}

\begin{figure}
\plotfiddle{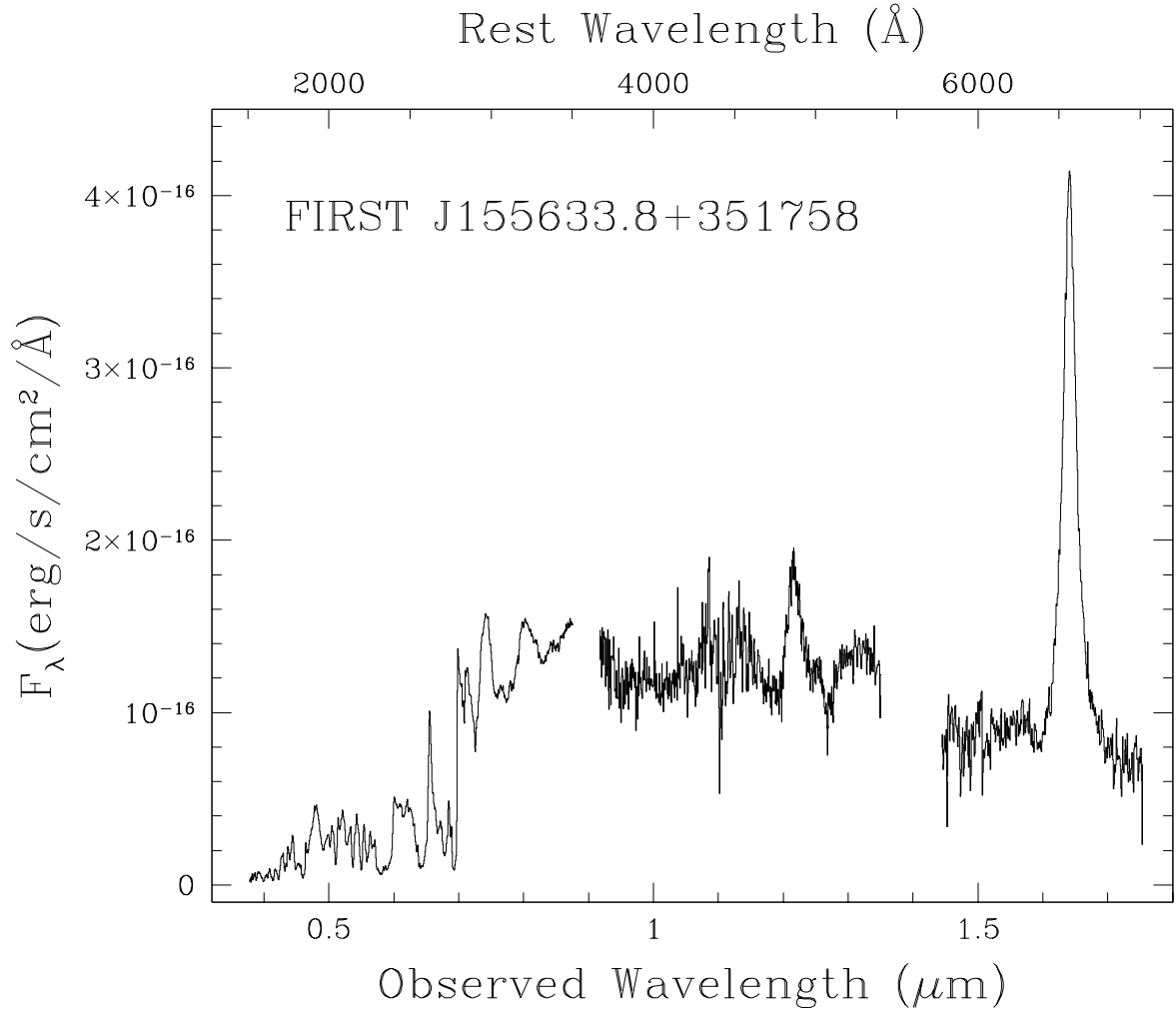}{6in}{0}{90}{90}{-250}{-110}
\figcaption{The optical and near-infrared spectrum of F1556+3517 
constructed by combining the total light Keck spectrum of 
Brotherton \etal (1997) and the KPNO 4-m data presented in this 
paper.  
\label{spectrum}
}
\end{figure}

The near-infrared spectroscopic data were combined with the optical 
spectrum of Brotherton et al.\ (1997), as shown in Figure~\ref{spectrum}.
The apparent discontinuity in the continuum between the 
$J$ and $H$-band observations is consistent with the 
continuum shape of other low-ionization BALQSOs 
which show broad, unexplained features redward of H$\beta$ in their 
\FeII-subtracted spectra 
(e.g., the MgII BALQSO 07598+6508; Boroson \& Meyers 1992). 
The optical spectrum is shown at the flux scaling adopted by 
Brotherton et al.\ (1997). 
Since the correction is small, no correction was performed for the Galactic 
extinction toward the object ($A_V=0.05$).

\begin{figure}
\plotfiddle{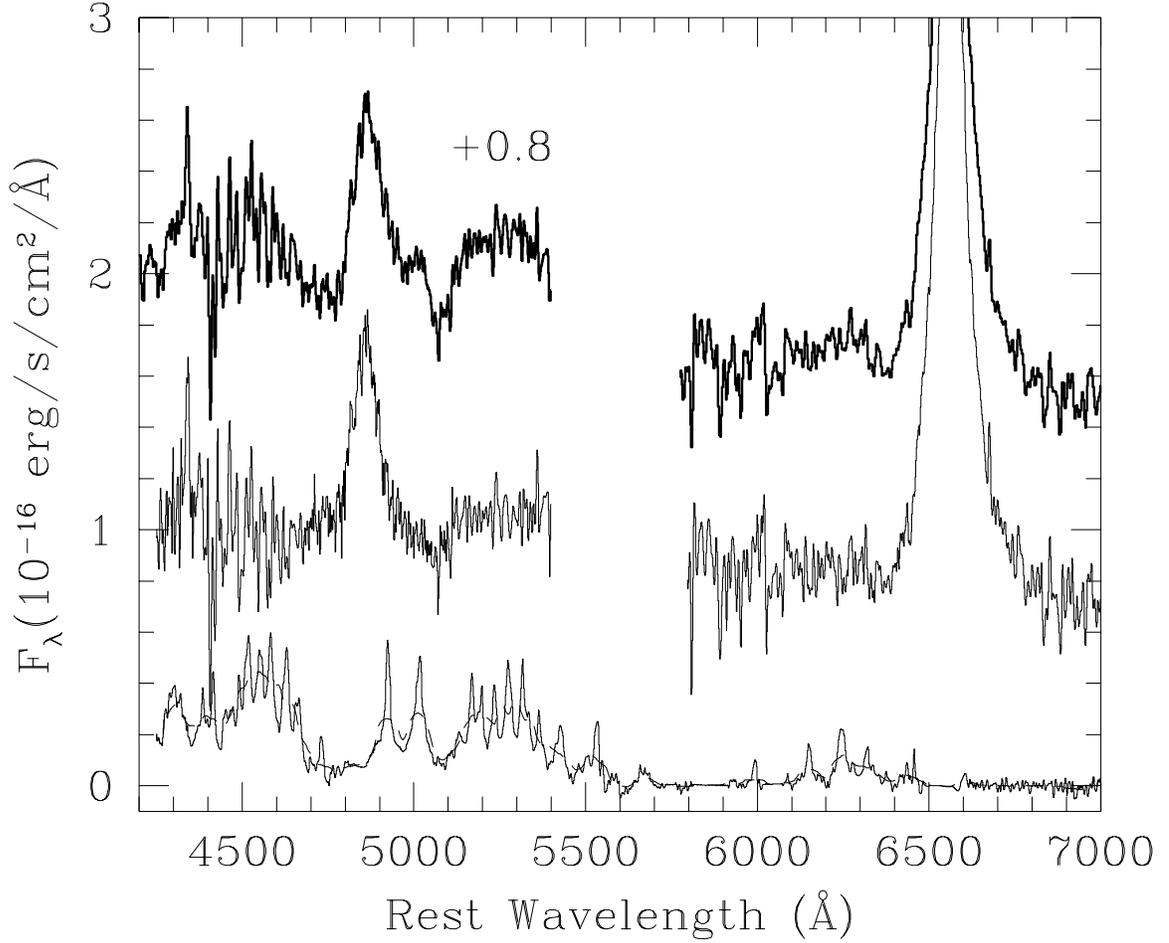}{6in}{0}{90}{90}{-250}{-110}
\figcaption{
The middle spectrum shows the F1556+3517 spectrum after subtraction 
of the smoothed FeII emission template of Boroson \& Green (1992). 
The original spectrum is shown for comparison as the upper 
solid line (offset by $0.8\times 10^{-16}$ erg/s/cm$^2$/\AA). 
The FeII emission template (bottom solid line) was smoothed 
by a gaussian of 4900 km\,s$^{-1}$ width (bottom dashed line). 
\label{nirspec} } 
\end{figure}

The observed continuum spectral energy distribution is red and 
strong emission lines of H$\alpha$ and H$\beta$ are readily apparent. 
A closer look at the near-IR region of the spectrum (Figure~\ref{nirspec})
reveals that \FeII\ emission is present near H$\beta$.  
In order to estimate the continuum shape in this region and to measure 
emission line properties, an \FeII\ emission template 
(Boroson \& Green 1992) was convolved with a gaussian with a width of 80\AA\ 
(i.e., $4900 \kms$, comparable to the width of the Balmer lines), scaled, 
and subtracted from the observed spectrum.  
The central wavelengths, fluxes, and widths of the emission lines, 
as measured from the 
\FeII-subtracted spectrum, are tabulated in Table~\ref{emdata}.
The QSO redshift,
as determined from the mean redshift of the H$\alpha$ and H$\beta$ lines, 
is $z_{em}=1.5008\pm 0.0007,$ where the error represents half the 
difference between the redshifts implied by the two lines.
Thus, the redshift is larger than that determined by Becker et al.\ (1997) 
for this object ($z=1.48$) from broad metal absorption features 
in the rest-frame UV and is in closer agreement with the redshift of 
$z=1.497\pm 0.001$ estimated by Brotherton \etal (1997) 
from a comparison of weak emission features in the rest-frame UV spectrum 
of F1556+3517 and a composite QSO spectrum.
Given that the presence of BAL absorption makes it difficult to locate 
the systemic velocity of metal (absorption and emission) lines, 
our redshift estimate is 
likely to be the most robust of the three estimates.
Thus, the broad metal absorption features are 
blue shifted with respect to the systemic velocity of the object. 

The ratio of the H$\alpha$ and H$\beta$ line fluxes is large 
($F_{\rm H\alpha}/F_{\rm H\beta} \approx 5.8$), 
much above the canonical case B value of 2.87 and typical values for QSOs
($\sim 3.5$; Rudy 1984), but well within the 
range of observed values for QSOs 
(H$\alpha$/H$\beta$=3--10; Baker 1997), 
and less extreme than the values measured by Egami et al.\ (1996)
for Hawaii~167 and 0059-2735 (13 and 7.6, respectively). 
The \Ha\ and \Hb\ line profiles are consistent given the 
uncertainties associated with the \FeII\ subtraction.
The H$\beta$ line is well fit with a single gaussian of
$\approx 5200$\kms\ FWHM.  The H$\alpha$ line, measured at 
higher signal-to-noise ratio, is better fit as a sum of two gaussians 
of 2800\kms\ and 6900\kms\ FWHM, respectively,
which has a combined FWHM of 4100\kms.
The \FeII\ emission contributes 18\% of the flux in the $J$ band
but only 2\% of the $H$ band flux.
The \FeII\ and H$\beta$ emission together contribute 30\% of
the $J$ band flux, while the H$\alpha$ line contributes 69\%
of the $H$ band flux.
The ratio of the \FeII-subtracted continua at \Ha\ and \Hb\ 
for F1556+3517 ($R_{\nu c}=1.54$) 
is similar to the ratios reported by Egami et al.\ (1996) 
for Hawaii~167 ($R_{\nu c}=1.32$)  
and 0059-2735 ($R_{\nu c}=1.67$).

Neither the H$\delta$, H$\gamma$, 
nor strong narrow emission lines were detected. 
The feature near $\lambda$4340\AA\ is much narrower (830 \kms)
than the \Ha\ and \Hb\ lines and is probably not H$\gamma$ 
emission associated with an AGN.
The 1-$\sigma$ upper limits to the strengths of 
H$\delta$, H$\gamma$, \OIII$\lambda$5007, and \OII$\lambda$3727, 
are 
$1.1\times 10^{-16} \ergpscm2$,
$1.4\times 10^{-16} \ergpscm2$,
$5.1\times 10^{-17} \ergpscm2$, and 
$1.5\times 10^{-16} \ergpscm2$.
The non-detection of narrow emission lines is in accord with previous
results for other low-ionization BALQSOs (e.g., Boroson \& Meyers 1992).
We do not detect the 4000\AA\ break 
($F_\lambda$[4050-4250\AA]/$F _\lambda$[3750-3950\AA]=1.03$\pm$0.02) 
in the F1556+3517 system (Figure~\ref{redstars}), 
unlike the situation for Hawaii~167 reported by Egami et al.\ 
(1996; $F_\lambda$[4000\AA]/$F_\lambda$[3727\AA]$\simeq 1.21$).

\section{Discussion}

\subsection{Extinction to the AGN}

The strength and width of the Balmer lines indicates that a large fraction of 
the rest-frame optical light arises from an AGN.  The polarization 
results reported for the rest-frame UV (Brotherton et al.\ 1997) 
indicate that photons reach us 
through both direct and scattered lines-of-sight. 
If we assume that the AGN is the sole source of rest-frame optical 
photons and that it is viewed through both a direct line-of-sight and 
a scattered line-of-sight, both the observed continuum slope and 
Balmer decrement, when compared with intrinsic values, allow us to 
place independent contraints on the dust column density along the direct 
line-of-sight.  
Note, however, that the unpolarized absorption troughs in the 
rest-frame UV may indicate the contribution of an additional source 
of continuum emission (e.g., a starburst).   If there is an additional 
continuum source, the Balmer lines would provide a better constraint on 
the extinction. 
Thus, the light we observe 
$$ F_\nu^{\rm obs} = F_d e^{-\tau_d} + F_s e^{-\tau_s} + F_{\rm trough}$$ 
may have contributions from an extincted, direct line of sight 
($F_d e^{-\tau_d}$), 
a typically less extincted, scattered line of sight 
($F_s e^{-\tau_s}$), 
and a third, unscattered source of continuum photons that fills in 
the unpolarized trough bottoms 
($F_{\rm trough}$).

The broad absorption and emission features imprinted on  
the scattered light spectrum of F1556+3517 
(Brotherton et al. 1997, figure 2, bottom panel) 
constrain the scattering geometry. 
The scattered light spectrum 
could be produced in a situation in which an electron scattering cloud 
is located within the BLR and the BAL region (cf. Ogle 1997, figure 2), 
but, unconventionally, {\it both} the direct and scattered lines of sight 
pass through the BLR and BAL region. 
Alternatively, the scattered light spectrum could indicate that 
the scatterer is located {\it beyond} the BLR and BAL region, and 
BAL-absorbed light from the QSO reaches the observer through both 
direct and scattered paths.  In the former case, the difference in 
the extinction between the two paths could reflect a spatial variation 
in the dust content of the BLR and/or BAL region.  In either case, 
the difference in extinction could also be due to a spatial variation 
in the dust content of the region beyond the BLR and BAL region. 

We account for either possibility, assuming 
that the scattered and direct components have the same 
intrinsic spectrum observed through different extinction paths.  
For simplicity, we assume 
that the scattering process is independent of wavelength, 
as in the case of electron scattering (e.g., Ogle 1997), and therefore  
$$ F_\nu^{\rm obs} = F_d^\prime\ (e^{-\Delta\tau} + f) + F_{\rm trough}$$ 
where $F_d^\prime = F_d\ e^{-\tau_s},$
$f=F_s/F_d,$ and $\Delta\tau=\tau_d - \tau_s$
represents the difference in extinction between the two paths. 

\subsubsection {Reddened and Scattered QSO Light}
We can consider first the constraints placed by the rest-frame optical 
spectrum, which may be dominated by the contribution from the direct 
line of sight, i.e., 
$ F_\nu^{\rm obs} \simeq F_d e^{-\tau_d}.$ 
For both Galactic (Savage \& Mathis 1979) and 
SMC (Bouchet et al.\ 1985) type dust, 
the observed Balmer decrement corresponds to a minimum 
$E(B-V)\le 0.7$ for intrinsic decrement values of 
$3-5.8$. 
These reddenings are comparable to the reddening implied by 
the observed continuum slope between \Ha\ and \Hb. 
Modeling the QSO continuum as a power-law,  
$F_d \propto \nu^{-\alpha}$, 
and fitting the observed continuum slope between \Ha\ and \Hb, 
for intrinsic QSO spectra of $\alpha$=($-$1, 0, 1), we find  
$E(B-V)\simeq$ (0.75, 0.46, 0.14) which correspond to intrinsic 
decrements of $\simeq$ (3, 3.5, 5).
Of these, the two higher extinction models predict a UV continuum 
level that is suppressed more strongly than is observed, while the 
lowest extinction model requires a highly unusual, very red 
intrinsic QSO spectrum (e.g., Elvis et al.\ 1994). 
Although these simple models provide a rough range for the extinction
in the system, a more realistic model must include a scattered
(and presumably less reddened) component, which is known to be present 
in the system (Brotherton et al.\ 1997). 

\begin{figure}
\plotfiddle{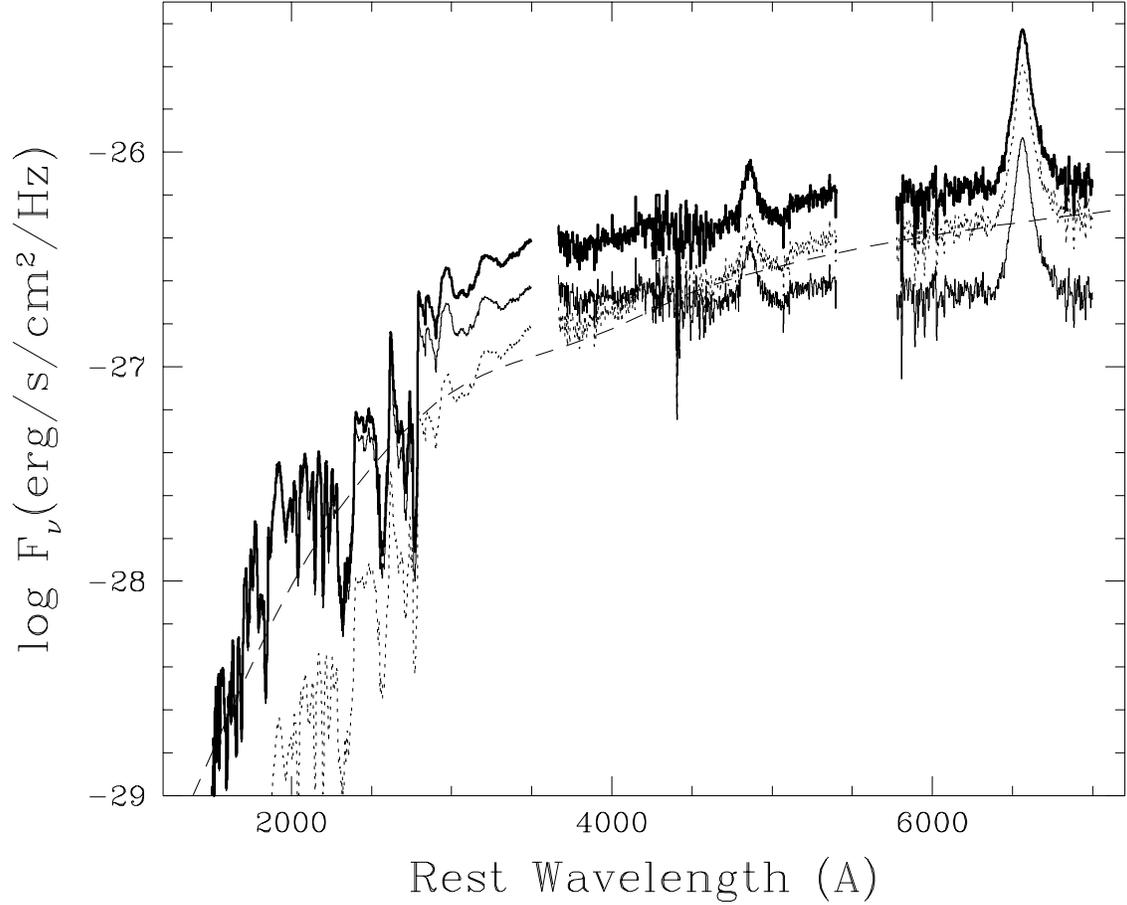}{6in}{0}{90}{90}{-250}{-110}
\figcaption{The \FeII-subtracted OUV spectrum of 1556+3517 
(heavy solid line) decomposed into a scattered contribution that 
dominates the flux at 2000\AA\   
(light solid line) and a direct contribution that experiences a 
differential extinction of $E(B-V)=0.5$ (dotted line). 
The intrinsic QSO continuum is flat, $F_\nu \approx$ constant as 
indicated by the scattered component.  
The direct component is well-fit by a $F_\nu \propto \lambda^0$  
continuum that is reddened by a total extinction of $E(B-V)=0.6$ 
(dashed line).  The additional $E(B-V)=0.1$ is needed to better 
approximate the downturn of the continuum in the rest frame UV.
\label{redscatqso} } 
\end{figure}

To explore this possibility, we consider 
a situation in which the light at 2000\AA\ is dominated by the 
scattered component, as is suggested by the high polarization of 
this spectral region (13\%; Brotherton et al.\ 1997).  
Figure~\ref{redscatqso} shows the contributions from the direct component 
$F_\nu^{\rm obs} e^{-\Delta\tau}/(e^{-\Delta\tau} + f)$ (dotted curve) 
and the scattered component $f F_\nu^{\rm obs}/(e^{-\Delta\tau} + f)$ 
(light solid curve) where $f=0.15$ and 
the difference in extinction between the direct and 
scattered lines of sight is $E(B-V)=0.5.$ 
In this figure $F_\nu^{\rm obs}$ is the \FeII-subtracted spectrum 
(heavy solid curve). 
The scattered flux is $\approx$10 times the direct flux at 2000\AA, 
the contributions are comparable at 4000\AA,
and the intrinsic QSO spectrum is characterized by $\alpha\sim 0$ 
and a Balmer decrement of $\sim 4.3$.
For comparison, we also show the continuum shape of an 
$\alpha=0$ spectrum that is extincted by the differential extinction 
of $E(B-V)=0.5$ suffered by the direct line of sight as well as 
an additional $\tau_s$ of $E(B-V)=0.1$ suffered by both lines of sight 
(dashed line).  
The non-zero value of $\tau_s$ was chosen 
to better match the shape of the UV continuum 
of the scattered component (light solid curve) at wavelengths 
$\lesssim$2900\AA.  

The intrinsic spectral slope $\alpha=0$ and decrement of 4.3 is in good 
agreement with 
the observed correlation between Balmer decrements and optical 
slopes in radio selected quasar samples.
From optical spectroscopy of a complete sample of low-frequency 
selected radio quasars (the Molonglo quasar sample), 
Baker (1997) measured Balmer decrements of $3.7-10$ and found 
that larger Balmer decrements are strongly correlated with 
steeper optical slopes, $F_\nu \propto \nu^{-\alpha}$ over the 
range $\alpha=0-2.5$.
This correlation was interpreted as evidence that all quasars 
have similar intrinsic optical slopes ($\alpha \simeq 0$) and 
Balmer decrements ($\approx 3.7$ with a spread $\pm0.5$) and that 
the range of observed values is primarily due to varying extinction 
along the line of sight to the AGN. 
Considering the two extinction estimates given above (extincted 
direct component only; extincted direct and scattered components) 
as limiting cases, it appears that 
F1556+3517 is low to moderately reddened, 
$E(B-V)=0.15-0.6$, with the actual value 
depending on the magnitude of the scattered contribution. 
Note that for a bluer intrinsic QSO continuum ($\alpha < 0$), the 
total reddening to the QSO would be larger than estimated here.

\subsubsection{Possible Starlight Contribution}
The third, unscattered component $F_{\rm trough}$ has been
ignored in the above discussion.  
This is appropriate if $F_{\rm
trough}$ has a flat spectral shape ($f_\nu\propto \nu^0$), 
as in the case of a young stellar 
population, or if the residual trough intensity is the result of partial 
covering of a (flat) QSO spectrum.  
Given the low flux density in the
UV troughs (10\% of the flux near {\ion{C}{3}]} $\lambda$1909; 
Brotherton et al.\ 1997), the contribution to the total light spectrum 
at longer
wavelengths would be negligible in this case.  However, this would not
be appropriate if $F_{\rm trough}$ rises rapidly in the red, as might 
be the case for an older or reddened stellar population.  
Brotherton et al.\ (1997) suggested that the emission in the unpolarized 
trough bottoms in the rest-frame UV spectrum of F1556+3517 
is due to a stellar component, e.g., 
as seen in Markarian 231 (Smith et al. 1995), a red quasar with a 
known starburst population. 
The rest-frame UV and optical spectra allow us to test this hypothesis.

\begin{figure}
\plotfiddle{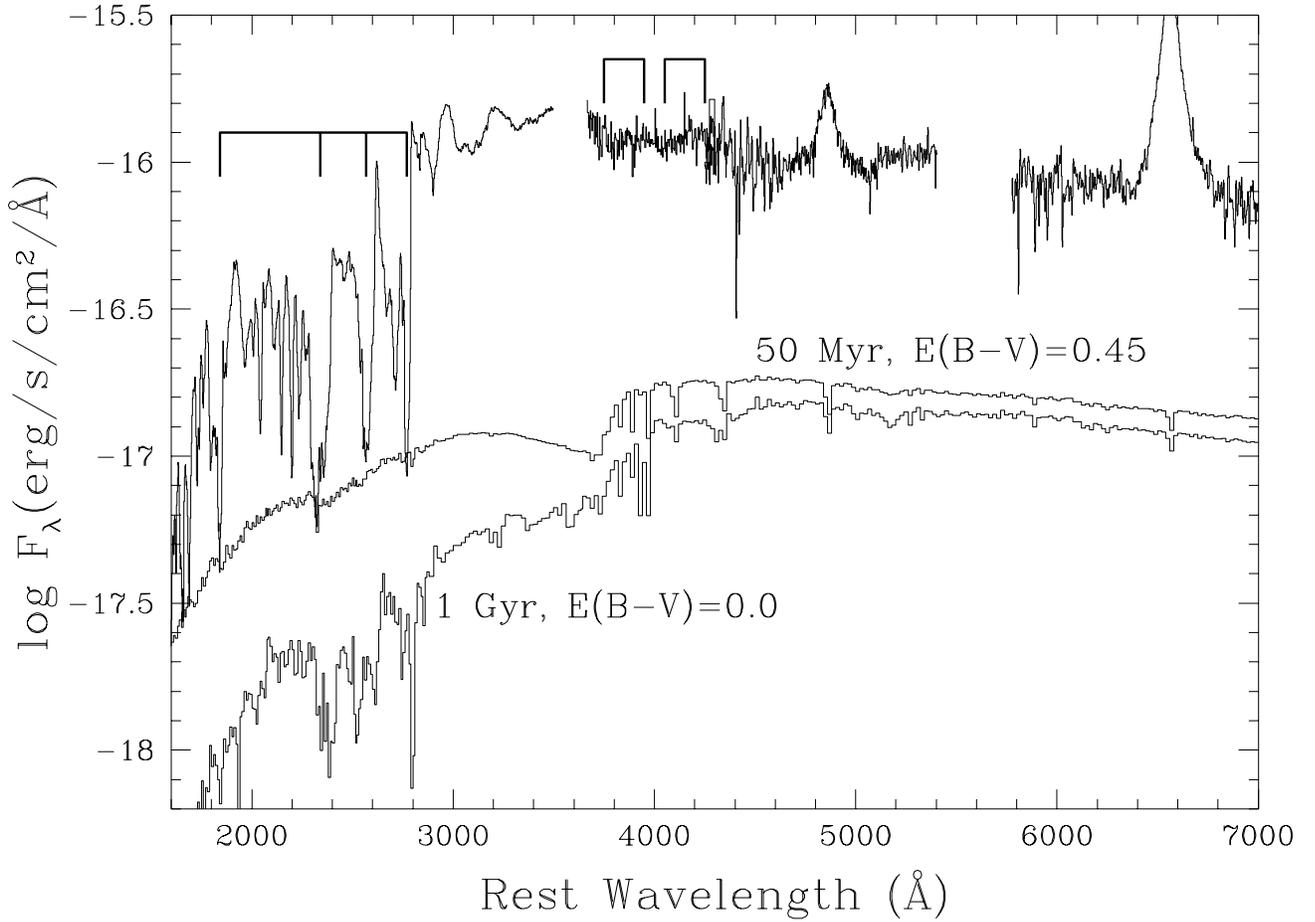}{6in}{0}{90}{90}{-250}{-110}
\figcaption{The spectral shape defined by the four trough bottoms 
at $\lambda<3000$\AA\ (left horizontal bar)
in the rest frame UV spectrum of F1556+3517 (top spectrum) 
is well fit by a 50~Myr old stellar population extincted by 
$E(B-V)=0.45$ (middle curve). 
An older, unreddened 1 Gyr population has a similar spectral shape 
in the rest frame UV (bottom curve), but because of its stronger 
4000\AA\ break (as defined by the wavelength regions indicated) 
can contribute no more than $\sim 20$\% of the trough light. 
\label{redstars} } 
\end{figure}

The four unpolarized low-ionization troughs located at rest-frame 
wavelengths 1860~\AA\ to 2800~\AA\ (Figure~\ref{redstars}) and our limit on the 
4000~\AA\ break constrain the possible starlight contribution to 
the spectrum. 
For example, 
the spectral shape defined by the four trough bottoms is fairly well 
fit by an unreddened 1 Gyr old population from the 1996 version of the 
Bruzual \& Charlot models (single burst, Salpeter IMF from $0.1-125\Msun$, 
solar metallicity; Bruzual \& Charlot 1993).  
However, it overpredicts the 4000\AA\ break and 
contributes $\sim$70\%  of the continuum at \Ha\ 
which implies that the QSO rest-frame \Ha\ equivalent width is 
improbably large ($\sim$1500\AA).
This old population is, therefore, unlikely to account for a 
significant fraction of the residual intensity in the troughs. 
To reduce the 4000\AA\ break strength of the composite spectrum to
within 1-$\sigma$ of the measured value requires that the 1 Gyr old
stellar population contribute $<$ 15\% of the flux at 4000\AA\ and
$\lesssim$ 20\% of the flux in the trough (Figure~\ref{redstars}, 
bottom curve).  
Assuming $H_0=80\hnotunit$ and $q_0=0.5,$ the stellar mass of this population 
is $< 3.3\times 10^{11}\Msun$. 

For comparison, we can also consider the limit placed on a possible 
starlight contribution from a younger stellar population. 
If we model the trough light as arising entirely from a 
50 Myr old population, we find that the spectral shape defined 
by the trough bottoms requires an extinction of $E(B-V)=0.45$ for 
the SMC extinction law (Figure~\ref{redstars}, middle curve).  
Since the intrinsic spectral energy distribution of a 50~Myr old population 
is flatter than that of a 1~Gyr old population, scaling the reddened 
model spectrum to match the flux density in the troughs implies a 
reasonable \Ha\ equivalent width.  
Although the younger population can contribute a greater 
fraction of the light at $\sim 2000$ \AA\ compared to the 1 Gyr 
population, the upper limit on the mass is lower in this case
($1.8\times 10^{11}\Msun$) due to the much smaller mass-to-light 
ratio of the younger population. 
Given these considerations, we conclude that 
if starlight contributes some fraction of the trough light, 
the flux contribution is small and has a negligible
impact on the extinction estimates presented in the previous sections.
We return in section 4.5 to a discussion of 
the possible stellar mass of the F1556+3517 system 
and its implications for the evolutionary history of QSOs.

\subsubsection{Comparison with Previous Extinction Estimates}

There have been a number of previous estimates of the reddening toward
F1556+3517, each employing a variety of techniques and assumptions.
Our analysis is more robust than previous estimates in that it 
uses the continuum slope and Balmer decrement as independent constraints   
and includes the contribution from a scattering component.  
Nevertheless, our extinction estimate is, perhaps fortuitously, in 
surprising general agreement with previous estimates. 
For example, from near-infrared photometry of F1556+3517,  
Hall \etal\ (1997) used the $B-K$ color of the object ($B-K=6.57$) to 
infer the extinction in the system.  They argued that since an intrinsic 
$B-K = 3$ is typical of QSOs at the redshift of F1556+3517 ($z=1.5$), 
$E(B-V) = 0.63$ assuming SMC-type extinction.  In retrospect, this 
analysis suffers from the assumption that F1556+3517 is a typical 
QSO when in fact strong BAL absorption depresses the flux shortward 
of $2800$\AA\ (rest frame), 
creating the artificial impression of a larger extinction. 
For example, the dereddened spectrum of F1556+3517
(section 4.1.2; $E(B-V)=0.5$) has a $B-R$ color of 2.25, 
whereas the mean QSO spectrum from the FIRST Bright Quasar Survey 
(Brotherton et al.\ 2000), 
when placed at the redshift of F1556+3517, has $B-R=0.55$.

In another study, Clavel (1998) used 
4--16 $\micron$ narrow-band images of F1556+3517 
obtained with ISOCAM to estimate the extinction.  
Comparing the measured infrared fluxes to published optical fluxes yielded 
an optical-IR spectral index of $\alpha = -2.$  
Assuming an intrinsic slope 
of $\alpha = -0.5$\ as likely and assuming a Galactic 
extinction law, Clavel determined $A_V = 1.6$.  
Whether observations extending over such a large range of wavelengths 
could be adequately modeled by screen extinction is a concern, 
since the flux at thermal infrared wavelengths may be affected by 
the reemission of extincted, reprocessed radiation, and the 
flux contribution at optical wavelengths is now known to have
a significant contribution from scattered light. 

Brotherton et al.\ (1997) estimated a higher extinction for 
F1556+3517 from the rest-frame UV spectrum ($E(B-V)=0.8$), 
due in part to their assumption of a bluer intrinsic QSO 
continuum slope ($\alpha = -0.3$) than is assumed here and the 
assumption that 
there are true continuum points in the spectrum at $\sim 1900$\AA.  
Since the location of the true continuum is difficult to determine 
from the rest-frame UV spectrum, which is contaminated by strong BAL 
absorption, the constraints placed by the present 
study are more robust.  We confirm that in a model 
in which the flux at 2000\AA\ is dominated by the scattering component, 
the extinction could be as high as $E(B-V)=0.8$ if the 
intrinsic continuum slope is $\alpha \simeq -0.3$. 
In this case, the scattered and direct components are comparable at 
4000\AA, the direct component contributes $\sim 80$\% of the flux 
at \Ha, and the intrinsic decrement is 3.4.

\subsection{The Radio Loudness of F1556+3517}

BALQSOs in general are radio quiet objects: the ratio of their radio to
optical continuum emission, generally parametrized by $\log
R^*\equiv\log (F_{5\rm GHz}/F_{2500})$ (where $F_{5\rm GHz}$ and
$F_{2500}$ are the 5~GHz and 2500\AA\ rest-frame $k$-corrected flux
densities), is small, with typical values of $\log R^*< 1.0$ (Stocke et
al. 1992).  In comparison, Becker et al.  (1997) report that F1556+3517
is unusually radio loud: calculating $\log R$ directly from the optical
blue magnitude estimated from the POSS-I plates, Becker et al. (1997)
found that F1556+3517 has $\log R\approx3.18$. Given our present
estimate of the extinction to the AGN, we can reexamine the issue of
radio loudness of F1556+3517 and compare it to other BAL and non-BAL
QSOs.

We calculated $\log R^*$ after correcting the optical magnitude for
both reddening and redshift. The observed flux density as measured from
the spectrum at $2500(1+z)$\AA\ is $5.6\times10^{-28}\,{\rm erg\,
s^{-1}\, cm^{-2}\, Hz^{-1}}$. Correcting for extinction using an
SMC-like extinction curve an $E(B-V)=0.5$ results in $\log R^*\approx
0.9$.  The main difference between our current estimate of $\log R^*$
and that presented by Becker et al. (1997) is the large extinction
correction: for our estimated $E(B-V)=0.5$, the corresponding
correction in the $B$ band (rest frame 1760\AA) for an SMC extinction
curve is $A_{1760}\approx 5.4$ mag.  The extinction corrected value of
$\log R^*$ derived here places F1556+3517 at the upper end of the $\log R^*$
distribution of known BAL QSOs,  at the borderline between radio loud
and radio quiet objects, and roughly near the peak of the distribution
for optically selected QSOs (\eg, Stocke et al. 1992;
see also Brotherton et al.\ 1998; Becker et al.\ 2000)

\subsection{Comparison to Hawaii~167 and 0059-2735}
The observational situation for F1556+3517 is superficially 
similar to that of Hawaii~167 and 0059-2735.  
As noted in section 3, the ratio of the continua at \Ha\ and \Hb\ 
is similar for all three objects, indicating modest reddening, but  
the Balmer decrement of Hawaii 167 is much larger 
than the observed decrements of both 0059-2735 and F1556+3517, 
indicating substantial reddening.
This discrepancy led Egami et al.\ (1994) to conclude that, at least 
in the case of Hawaii~167, the continuum shape is not a reliable indicator 
of the extinction to the AGN.  For Hawaii~167, the extinction implied 
by the Balmer decrement was large enough to extinguish the rest-frame 
UV continuum of the AGN.  As a result, the authors concluded that the 
UV continuum originated elsewhere, most probably in a starburst, 
which the UV spectrum (Cowie et al.\ 1994) resembles.
In the case of F1556+3517, the UV spectrum does not resemble that 
of a starburst, 
and the more modest Balmer decrement leaves 
open the possibility that both the QSO continuum slope and the 
Balmer decrement probe the same, low to moderate, dust column density. 
In this interpretation, the extreme (observed) $B-K$ color of F1556+3517, 
compared to the less extreme color of Hawaii 167,  
is due to the combined absence of stellar UV continuum light and 
the suppression of the UV continuum of the AGN by strong 
BAL absorption, rather than due to a significant column density of 
intervening dust.

\subsection{Relation Between Gas and Dust Column Densities}
It is interesting to compare the derived dust absorption column density
with estimates for the column densities in BALs and the typical X-ray
absorption column densities for BALQSOs.  BALQSOs have been found to be
weaker X-ray sources than non-BALQSOs with otherwise similar properties
(Green \& Mathur 1996; Gallagher et al. 1999; Brinkmann et al. 1999), a
result which is attributed, a least in part, to absorption of the X-ray
flux by a substantial intervening gas column density ($N_H>10^{22} {\rm
cm}^{-2}$ to $5\times 10^{23} {\rm cm}^{-2}$).  Similarly large column
densities are also implied in analyses of BAL features.  For example,
Hamann (1998) has shown that the detection of broad \ion{P}{5}
absorption in the BALQSO PG1254+047 and the strengths of the BALs in
that system are consistent with solar abundance for the BALs if many of
the lines are severely saturated and the total hydrogen column density
is high ($N_H > 10^{22} {\rm cm}^{-2}$).  Such large column densities
are estimated to provide sufficient optical depth in a large enough
number of lines to radiatively drive the BAL outflow.

Evidence for a large intervening gas column density in F1556+3517
is available from a serendipitous observation of F1556+3517 with 
the ROSAT PSPC; F1556+3517 is not detected in an 8 ksec exposure. 
If we estimate the intrinsic X-ray flux of F1556+3517 using the 
radio$-$X-ray luminosity correlation of Siebert et al.\ (1996) 
and assuming an X-ray photon index of 1.7 (cf., the radio-loud QSO 
B2\,1225+31 at z$=$2.2, Buehler et al.\ 1995),  
the ROSAT non-detection implies an intrinsic $N_{\rm H} \geq
3\times10^{23}$\,cm$^{-2}$.  
In comparison, the dust absorption column density averaged along 
lines of sight to the rest-frame UV/optical continuum emitting region 
of F1556+3517 is $A_V \simeq 1.5,$ which corresponds to a much lower gas 
column density of $N_H \simeq 3\times 10^{21} {\rm cm}^{-2},$ 
given the standard gas-to-dust ratio for the Galaxy, 
or $N_H \simeq 3\times 10^{22} {\rm cm}^{-2}$ 
with the gas-to-dust ratio for the SMC. 
If this measurement probes the equivalent line-of-sight as the X-ray 
measurements, then, our measured 
dust column density could be either intermixed with the X-ray/BAL absorbing gas 
or located further along the line-of-sight to the observer, i.e., outside 
the BAL region.
In either case, the much lower dust absorption column densities that 
we measure may indicate that the X-ray/BAL absorbing gas is quite dust-poor, 
with a gas-to-dust ratio lower even than that of the SMC. 

The absence of significant dust in the broad line region (BLR) is consistent 
with the results of Laor \& Draine (1993) who have argued that the BLR is 
too close to the active nucleus for much dust to survive in the BLR.  
Indeed, as argued by Netzer \& Laor (1993), the radial extent of the BLR may 
be defined by the dust sublimation radius.  
If the BAL region is located just beyond the BLR, dust may survive within it.
Scoville \& Norman (1995) have proposed that radiation pressure on dust 
supplied by circumstellar mass loss from evolved stars residing in a 
nuclear star cluster is responsible for the ejection of BAL outflows.  In 
their analysis, they assume that the BAL gas is metal rich, and therefore 
dust-rich, with a Galactic dust-to-gas ratio.  
If the BAL gas column density 
in F1556+3517 is at least that measured for PG1254+047, we find that 
the BAL region has a much lower dust-to-gas ratio, $<1/3$ the Galactic value.  
Since the rest-frame UV continuum of F1556+3517 is more
strongly suppressed by BAL absorption than is PG1254+047 (Steidel \&
Sargent 1991), the gas absorption column of F1556+3517 is likely to be
larger than that of PG1254+047 and the dust-to-gas ratio even lower than
estimated above.
Moreover, the possibility that the dust column we measure is not 
associated with the BAL also suggests that the actual dust-to-gas ratio 
is actually much less than the Galactic value.  
If most BAL regions are 
this dust poor, it may prove difficult to drive BAL outflows using radiation 
pressure on dust grains.  Resonance-line driven winds 
(e.g., Murray \& Chiang 1995) 
would then be the more likely possibility. 

As a caveat to the above discussion, 
we note that it is likely that the optical continuum and Balmer line 
emitting region is intrinsically much larger than the X-ray emitting region and, 
therefore, probes a wider range of lines-of-sight compared to observations made 
at X-ray wavelengths.  In this case, the lower column density may result  
from an average over predominantly low column density lines-of-sight to the 
optical continuum emission region compared to  
higher column density lines-of-sight to the X-ray emitting region.
In this case, determining the appropriate correction factor that accounts for the 
different lines of sight would be necessary before we can accurately compare the 
gas and dust column densities measured at different wavelengths.

\subsection{Stellar Content and Evolutionary State of F1556+3517}

Magorrian et al.\ (1998) have shown that the black hole masses of 
nearby galaxies 
are remarkably correlated with their estimated bulge masses over 4 
orders of magnitude, with $M_{\rm BH}/M_{\rm bulge}\simeq 0.005$;
more recent estimates suggest a smaller proportionality constant 
$M_{\rm BH}/M_{\rm bulge}\simeq 0.0017$ (Gebhardt et al.\ 2000).
If this implies a direct relation between the growth rates of central 
black holes and stellar bulges, this argues that accretion on parsec 
size scales is somehow related to accretion on kiloparsec size scales.
Silk \& Rees (1998) argue that since 
collapsing objects tend to form cuspy density profiles 
(e.g., Navarro \& Steinmetz 2000), the early formation of a massive 
central black hole is greatly favored over the early formation 
of a large number of stars distributed on galactic size scales. 
They also argue that a wind from the QSO that forms, if 
dynamically significant, i.e., possessing adequate force to 
reverse accretion onto the black hole, will tend to produce a 
correlation between black hole mass and galactic mass that is 
much like the observed correlation. 
Expanding on this hypothesis, Fabian (1999) argues more explicitly 
that the stellar spheroid and central black hole grow together 
by the coordinated accretion of gas on kiloparsec and sub-parsec 
size scales.  The presence of cold gas in the 
vicinity of the black hole is responsible for the heavy obscuration 
that is believed to characterize the early evolutionary history of 
QSOs.  In Fabian's picture, the powerful wind from the QSO 
is eventually able to sweep away the obscuration, 
revealing the QSO and terminating the growth of both the black hole and 
the stellar component of the galaxy.

Since F1556+3517 is one such dust-enshrouded QSO with a powerful outflow 
that may be viewed in an early evolutionary state, it is of interest to 
explore whether and how F1556+3517 fits into this picture.
As described in section 4.1.3, we estimated an upper limit to the stellar 
mass of the system by assuming that a maximal fraction of the light 
observed in the unpolarized UV troughs is due to stars. We found that 
if the stellar population is of intermediate age (1 Gyr), it can 
contribute no more than $3.3\times 10^{11}\Msun.$  If the stellar 
population is young (50 Myr), the maximal stellar mass is 
$1.8\times 10^{11}\Msun$.
Since the 1\arcsec\ and 1\farcs5 slit widths used in our observations 
correspond to linear dimensions of 5.5 kpc and 8.2 kpc  
($H_0=80\hnotunit$ and $q_0=0.5$),  
observations are likely to account for the entire bulge population 
in the system. 

We can estimate the black hole mass of the QSO from the dereddened 
direct component of the rest-frame 
$V$-band continuum flux.  Using the mean QSO SED from Elvis et al.\ (1994), 
which implies that $L_{\rm bol}/L_V=13.2,$ we obtain 
$L_{\rm bol}=1.8\times 10^{13}\Lsun$.
Assuming that the black hole radiates at a fraction $\eta$ of the 
Eddington luminosity, the implied black hole mass  is 
$M_{\rm BH}=5.7\times 10^8/\eta \Msun$. 
Thus, the ratio of the black hole and stellar masses is $0.0017/\eta$. 
If $\eta=1$, the ratio is the same as that given by the 
Magorrian relation. 
If $\eta$ is instead a more likely value 
($\eta < 0.1$; Osterbrock 1989; Nulsen \& Fabian 2000), 
then the black hole mass is much larger than would 
be implied by the stellar mass given the Magorrian relation.

As described by Laor (1998; see also Laor [2000] and references therein), 
the \Hb\ linewidth can be used to obtain an estimate of the black hole 
mass and $\eta$ based on the general assumptions that velocities in the 
broad line region are governed by gravity 
(i.e., $v^2 \simeq G M_{\rm BH}/R_{\rm BLR}$) 
and that the size of the BLR is set by the bolometric 
luminosity of the AGN. 
Laor (1998) has adopted a relation of the form
$R_{\rm BLR} = 0.086\,L_{46}^{1/2}$\,pc 
(where $L_{46}=L/10^{46}\ergps$ and $L$ is the bolometric luminosity), 
based on the reverberation mapping results of Kaspi et al.\ (1996)
and other theoretical considerations.  
Taken together, these two assumptions imply a relation between the 
black hole mass, characteristic BLR velocity, and AGN luminosity of 
$M_{\rm BH}=0.18\times 10^9 \Msun\,v_{3000}^2\,L_{46}^{1/2}$, 
and, therefore, $\eta= 0.44\,v_{3000}^{-2}\,L_{46}^{1/2}$.

Given the relative lack of specificity in the physics underlying these 
relations, one would expect 
that although the relations might describe general 
correlations between AGN properties, they are unlikely to predict 
black hole masses accurate to better than an order of magnitude. 
Surprisingly, if $v$ is taken to be the FWHM of the \Hb\ line and 
black hole masses are predicted for a sample of PG quasars with 
measured bulge luminosities, the relation between bulge luminosity and
black hole mass overlies the Magorrian relation (Laor 1998), a result 
which has been taken to imply that the scaling relation is accurate 
to a factor of $2-3$.
It is important to bear in mind, however, that only a direct measurement 
of $M_{\rm BH}$ (and therefore $\eta$) in these systems would confirm the 
quantitative accuracy of such scaling relations.
Therefore, we adopt a more cautious approach and regard the 
black hole mass predicted by such scaling relations as a speculative, 
but interesting possibility.

Using the above scaling relation and the measured \Hb\ width 
for F1556+3517, we find 
$\eta=0.40$ and $M_{\rm BH} = 1.4\times 10^9\Msun.$ 
If $\eta=0.40$ and the stellar mass is at its maximal allowed value  
($3.3\times 10^{11}\Msun$), the ratio of $M_{\rm BH}/M_{\rm bulge}=0.004$ 
puts F1556+3517 close to the mean Magorrian relation, a result 
similar to that found by Laor (1998) for the PG quasar sample.
In this case, the higher redshift of F1556+3517, compared to the 
PG quasar sample, implies that 
the Magorrian relation is already in place at $z=1.5.$ 
Moreover, at the maximal stellar mass the galaxy would already have 
formed most of its stars.  So if F1556+3517 eventually evolved 
to fall on the Magorrian relation, the evolution to the final black 
hole mass was already nearly complete at $z=1.5.$

The relatively large value of $\eta$ thus derived for F1556+3517 
is consistent with some of its other spectral properties.
The strong \FeII\ and weak \OIII\ emission
of F1556+3517 place it in a group of objects 
that share extreme values of the Boroson and Green `eigenvector 1'
(Boroson and Green 1992).
This group, which includes Narrow Line Seyfert 1s, low ionization BAL QSOs, 
and QSOs with weak soft X-ray emission (Brandt, Laor \& Wills 2000), 
share traits that were
postulated by Boroson and Green (1992) to imply large accretion rates.
Recent estimates by Laor (2000) based on the scaling relation described
above confirm that these strong FeII QSOs have large values of $\eta$.
The reason behind the high accretion luminosities (presumably driven by
a high accretion rate) in these AGN is unknown:  various authors have
speculated that it may indicate that these AGN are young, based on the
tendency for these sources to have strong associated absorption,
infrared excesses, and (in the case of the nearest examples) show
evidence of recent star burst activity (\eg, Mathur 2000; Becker et
al.\ 2000, and references therein). 

However, other considerations favor a higher $M_{\rm BH}/M_{\rm bulge}$. 
For example, more recent results reported 
by Kaspi et al.\ (2000) favor a steeper $R_{\rm BLR}(L)$ relation:  
$R_{\rm BLR}=0.17\,L_{46}^{0.7}$ pc (for $H_0=80\hnotunit$ $q_0=0.5$). 
With this relation, 
$M_{\rm BH}=0.34\times 10^9 \Msun\,v_{3000}^2\,L_{46}^{0.7}$, 
and $\eta= 0.23\,v_{3000}^{-2}\,L_{46}^{0.3}$.
Therefore, in the case of F1556+3517, 
$\eta=0.14$ and $M_{\rm BH} = 4.1\times 10^9\Msun.$
With this lower value of $\eta$ and a maximal stellar mass,  
$M_{\rm BH}/M_{\rm bulge}=0.012,$ significantly above the 
Magorrian relation. 

Note also that the black hole mass is estimated conservatively. 
The $L/L_{\rm Edd}$ of the active nuclei of local Seyferts 
are found to be as small as $\eta =10^{-4}$ (E. Agol, personal 
communication).
In addition, a maximal scattered fraction of the F1556+3517 spectrum, as
determined in section 4.1.1, has been excluded in the luminosity
estimate.  We have assumed that the black hole radiates isotropically,
a conservative assumption given the probable edge-on orientation of the
system.  The extinction correction we have applied ($A_V=1.55$) is
modest, and unlikely to be much smaller since intrinsic QSO continua 
are likely to be bluer rather than redder than we have assumed.  

In contrast, the stellar mass is likely to be overestimated.
We have assumed that a maximal fraction of the trough light is 
starlight due to a 1 Gyr population, whereas 
a fraction could be unpolarized light from the active nucleus, either 
due to partial covering of the nucleus or multiply scattered nuclear emission.
If we have overestimated the age of the stellar population, as is likely 
to be the case, the true stellar mass is even smaller.  
We have also assumed a Salpeter IMF, which has a more massive contribution 
from low mass stars than is measured for the stellar field population in 
the solar neighborhood.  
Thus, it appears likely that $M_{\rm BH}/M_{\rm bulge}$ is larger 
than estimated above.  Consequently, if F1556+3517 eventually evolved 
to fall on the 
Magorrian relation, then, as speculated by Silk and Rees (1998), the 
black hole mass developed more rapidly than the 
bulge mass, a massive black hole already having been in place at $z=1.5$. 
Whether this evolutionary history is anomalous or the norm among QSOs 
is an issue of considerable interest for the future.

\section{Conclusions}
Our near-IR spectrum of the unusual BALQSO F1556+3517, combined with the
existing optical spectrum, places new constraints on the dust and stellar
content of the system. These results suggest interesting implications
for the origin of QSO outflows and the accretion history of QSOs.

Both the Balmer decrement and the rest-frame UV-optical continuum slope
of F1556+3517 provide evidence for modest extinction in the system. By
modelling the rest-frame UV-optical spectrum as a combination of extincted
direct and scattered light components, we find that the differential
extinction between the two lines of sight is $E(B-V)\approx 0.5$. The
total extinction to the active nucleus may be higher if the intrinsic
spectrum of the QSO is bluer than we have assumed (i.e., if $\alpha<0$;
$F_\nu\propto\nu^{-\alpha}$).  Since the extinction is modest, the
very red $B-K$ color of F1556+3517 is due to the suppression of the
rest-frame UV continuum by strong BAL absorption rather than due to
a large column density of dust along the line of sight.  For a typical
Galactic dust-to-gas ratio, the implied gas column density along the line
of sight to the broad line region is much less than has been inferred for
F1556+3517 on the basis of its weak X-ray flux.  This suggests that either
the broad emission and absorption line regions have a low dust-to-gas
ratio or that the rest-frame optical light encounters significantly
lower mean column density lines of sight than the X-ray emission.
If our results are due to a low dust-to-gas ratio, it would appear
difficult to drive BAL outflows with radiation pressure on dust grains,
and resonant line driven winds are a more attractive possibility.

In contrast to the result obtained by Egami et al. (1996) for 
the red quasar Hawaii 167, the 4000\AA\ break is not detected in the system, 
implying that if the system has a significant stellar population, 
the population is young ($< 1$ Gyr). 
Assuming the maximal stellar contribution allowed by the spectrum, 
and using the rest-frame $V$-band continuum
to estimate the black hole mass, we find that 
if $L/L_{\rm Edd}$ is large ($\sim 1$), the ratio 
$M_{\rm BH}/M_{\rm bulge}$ is comparable to that of the Magorrian relation.  
In this case, the Magorrian relation is already in place at $z=1.5$.
However, multiple factors favor a much larger $M_{\rm BH}/M_{\rm bulge}$ 
ratio.  In this case, if the Magorrian
relation characterizes the late history of QSOs, and the situation
observed for F1556+3517 is typical of the early evolutionary history of
QSOs, this result implies that central black hole masses develop more
rapidly than bulge masses.

\acknowledgements

We are very grateful to the staff of KPNO, particularly Dick Joyce and
Brett Huggard, who made these observations possible, and Paul Ho 
who assisted with the observations. We thank Paul
Martini and Pat Hall for providing us with their $K$-band image of
F1556+3517 in advance of publication.  We also thank Eric Agol for
communicating his results on the $L/L_{\rm edd}$ of Seyfert galaxies in
advance of publication.  We are also grateful to Ed Moran and Sally
Laurent-Muehleisen for assistance with the ROSAT data on F1556+3517; to
Gary Bower, Todd Boroson, Richard Green, and especially Ari Laor for
their comments on the manuscript; and to the anonymous referee whose
thoughtful comments improved the presentation of the paper.  Finally,
we thank Sperello di Serego Alighieri and the staff of the Osservatorio
di Arcetri for their kind and generous hospitality in Florence, where
an early draft of this paper was written.

\pagebreak

\centerline {\bf References}

\refindent Baker, J. C. 1997, \mnras, 286, 23

\refindent Baker, A.C., Carswell, R.F., Bailey, J.A., Espey, B.R., 
Smith, M.G.\ \& Ward, M.J.\ 1994, MNRAS, 270, 575

% Properties of Radio-selected BAL Quasars from the 
% FIRST Bright Quasar Survey:
\refindent Becker, R.H., White, R.L., Gregg, M.D., Brotherton, M.S., 
Laurent-Muehleisen, S.A., Arav, N. 2000, \apj, 538, 72 

\refindent Becker, R.\ H., Gregg, M.\ D., Hook, I.\ M., McMahon, R.\ G., White, R.\ L.\ \& Helfand, D.\ J.\ 1997, \apjl, 479, L93

\refindent Boroson, T.\ A.\ \& Green, R.\ F.\ 1992, \apjsupp, 80, 109

\refindent Boroson, T.\ A.\ \& Meyers, K.\ A.\ 1992, \apj, 397, 442

\refindent Bouchet, P., Lequeux J., Maurice, E., Pr\'evot, L., \& 
Pr\'evot-burnichon, M.\ L.\ 1985, AA, 149, 330 

\refindent Brandt, W.\ N., Laor, A.\ \& Wills, B.\ J.\ 2000, \apj, 528, 637

%\refindent Brinkmann, W., Morganti, R., Tadhunter, C. N., Danziger, I. J., Fosbury, R. 
%A. E. and di Serego Alighieri, S. 1996, \mnras, 279, 1331 

\refindent Brinkmann, W., Wang, T., Matsuoka, M. \& Yuan, W. 1999, \aap, 
345, 43 

\refindent Brotherton, M. S., Tran, H. D., van Breugel, W., Dey, A.\ \& 
Antonucci, R.\ 1997, \apjl, 487, L113

% Discovery of Radio-Loud BAL Quasars Using UV Excess and Deep 
% Radio Selection: 
\refindent Brotherton, M.S., van Breugel, W., Smith, R.J., Boyle, B.J., 
	Shanks, T., Croom, S.M., \& Miller, L. 1998, 
	\apj, 505, L7

\refindent Brotherton, M. S., Tran, H. D., Becker, R.H., Gregg, M.D., 
Laurent-Muehleisen, S.A., \& White, R.L. 2000, ApJ, submitted

\refindent Bruzual A., G. \& Charlot, S. 1993, \apj, 405, 538 

\refindent Buehler, P., Courvoisier, T. J.-L., Staubert, R., Brunner, H. \&
Lamer, G. 1995, \aap, 295, 309

\refindent Clavel, J. 1998, A\&A, 331, 853 

\refindent Cohen, M. H., Ogle, P.  M., Tran, H. D., Vermeulen, R. C., Miller, J.
S., Goodrich, R. W. \& Martel, A. R. 1995, \apjl, 448, L77

\refindent Cowie, L.\ L., Songaila, A., Hu, E.\ M., Egami, E., 
	Huang, J.\ S., Pickles, A.\ J., Ridgway, S.\ E., Wainscoat, R.\ J., 
	\& Weymann, R.\ J.\ 1994, ApJ, 432, L83

\refindent Egami, E., Iwamuro, F., Maihara, T., Oya, S.\ \& Cowie, L.\ L.\ 
1996, \aj, 112, 73

\refindent Elias, J.H., Frogel, J.A., Matthews, K., \& Neugebauer, G. 1982, 
AJ, 87, 1029

\refindent Elvis, M.\ et al.\ 1994, \apjs, 95, 1 

\refindent Fabian, A.C. 1999, MNRAS, 308, L39

\refindent Gallagher, S. C., Brandt, W.N., Sambruna, R.M., 
Mathur, S.\ \& Yamasaki, N.\ 1999, ApJ, 519, 549

\refindent Gebhardt, K. et al.\ 2000, astro-ph/0007123, \apjl, in press

\refindent Goodrich, R.W., \& Miller, J.S. 1995, ApJ, 448, L73

\refindent Green, P.J., \& Mathur, S. 1996, ApJ, 462, 637

\refindent Hall, P.\ B., Martini, P., DePoy, D.\ L.\ \& Gatley,
I.\ 1997, \apjl, 484, L17

\refindent Hamann, F.\ 1998, ApJ, 500, 798

\refindent Hazard, C., McMahon, R.\ G., Webb, J.\ K., \& Morton, D.\ C.\ 
1987, \apj, 323, 263

\refindent Hines, D. C., Schmidt, G. D., Smith, P. S., Cutri, R. M. \& 
Low, F. J.  1995, \apjl, 450, L1

\refindent Hines, D.\ H.\ \& Schmidt, G.\ D.\ 1997, in 
Mass Ejection from Active Galactic Nuclei, 
ed. N. Arav, I. Shlosman, \& R.\ J.\ Weymann (San Francisco: ASP), 59

\refindent Joyce, R.R.\ 1995, ``Cryogenic Spectrometer Users Manual'', (Tucson:NOAO)

\refindent Joyce, R.R., Fowler, A.M., \& Heim, G.B.\ 1994, Proc.\ SPIE, 2198, 725.

\refindent Kaspi, S., Smith, P.S., Maoz, D., Netzer, H., \& Jannuzi, B.T., 
	1996, ApJ, 471, L75

\refindent Kaspi, S., Smith, P.S., Netzer, H., Maoz, D., Jannuzi, B.T., 
	\& Giveon, U. 2000, \apj, 533, 631

\refindent Laor, A. \& Draine, B. T. 1993, \apj, 402, 441 

\refindent Laor, A., Jannuzi, B.\ T., Green, R.\ F.\ \& Boroson, T.\ A.\ 
1997, \apj, in press. 

\refindent Laor, A. 1998, \apj, 505, L83

\refindent Laor, A. 2000, astro-ph/0005144, to appear in New Astronomy 
	Reviews

\refindent Magorrian, J., Tremaine, S., Richstone, D., Bender, R., Bower, G., 
	Dressler, A., Faber, S.M., Gebhardt, K., Green, R., Grillmair, C., 
	Kormendy, J., \& Lauer, T. 1998, AJ, 115, 2285

\refindent Mathur, S.\ 2000, MNRAS, 314, L17

\refindent Murray, N., \& Chiang, J. 1995, ApJ, 454, L105

\refindent Navarro, J.F., \& Steinmetz, M. 2000, ApJ, 528, 607

\refindent Netzer, H. \& Laor, A. 1993, \apjl, 404, L51 

\refindent Nulsen, P.E.J., \& Fabian, A.C. 2000, MNRAS, 311, 346

\refindent Ogle, P.\ M.\ 1997, in  
Mass Ejection from Active Galactic Nuclei, 
ed. N. Arav, I. Shlosman, \& R.\ J.\ Weymann (San Francisco: ASP), 78

\refindent Ogle, P. M., Cohen, M. H., Miller, J. S., Tran, H. D., 
Goodrich, R. W. \& Martel, A. R. 1999, \apjs, 125, 1 

\refindent Osterbrock, D.E. 1989, 
	Astrophysics of Gaseous Nebulae and Active Galactic Nuclei, 
	(Mill Valley: University Science Books), p. 342

\refindent Rudy, R.J. 1984, ApJ, 284, 33

\refindent Sanders, D.B., Soifer, B.T., Elias, J.H., Neugebauer, G., 
	\& Matthews, K. 1988, ApJ, 328, L35

\refindent Savage, B.\ D.\ \& Mathis, J.\ S.\ 1979, ARAA, 17, 73

\refindent Schmidt, G. D. \& Hines, D. C. 1999, \apj, 512, 125 

\refindent Scoville, N., \& Norman, C. 1995, ApJ, 451, 510

\refindent Siebert, J., 
Brinkmann, W., Morganti, R., Tadhunter, C. N., Danziger, I. J., Fosbury, R. 
A. E. \& di Serego Alighieri, S. 1996, \mnras, 279, 1331
%The soft X-ray properties of a complete sample of radio sources.

\refindent Silk, J., \& Rees, M.J. 1998, A\&A, 331, L1

\refindent Smith, P. S., Schmidt, G. D., Allen, R. G. \&
Angel, J. R. P. 1995, \apj, 444, 146 

\refindent Sprayberry, D. \& Foltz, C. B. 1992, \apj, 390, 39 

\refindent Steidel, C.C., \& Sargent, W.L.W. 1991, ApJ, 382, 433

\refindent Turnshek, D.\ A., et al.\ 1994, ApJ, 428, 93

\refindent Turnshek, D.\ A., Monier, E.\ M., Sirola, C.\ J.\ \& Espey,
B.\ R.\ 1997, \apj, 476, 40

\refindent Voit, G.\ M., Weymann, R.\ J.\ \& Korista, K.\ T.\ 1993, \apj, 413, 95

%\refindent Webster, R.\ L., Francis, P.\ J., Peterson, B.\ A., 
%Drinkwater, M.\ J.\ \& Masci, F.\ J.\ 1995, \nature, 375, 469

\refindent Weymann, R.\ J., Morris, S.L., Foltz, C.B., \& Hewett, P.C. 1991, 
	ApJ, 373, 23

\refindent Weymann, R.\ J., Carswell, R.\ F., \& Smith, M.\ G.\ 1981, \araa, 19, 41

%\refindent Weymann, R.\ J., Williams, R.\ E., Peterson, B.\ M.\ \& Turnshek, 
%D.\ A.\ 1979, \apj, 234, 33

\medskip

%%%%%%%%%%%%%%%%%%%%%%%%%%%%%%%%%%%%%%%%%%%%%%%%%
%% TABLES                             %%%%%%%%%%%
%%%%%%%%%%%%%%%%%%%%%%%%%%%%%%%%%%%%%%%%%%%%%%%%%

% TABLE ONE: EMISSION LINE DATA
\begin{deluxetable}{lccccc}
\tablewidth{0pt}
\tablecaption{Spectral Features}
\tablehead{
	  \colhead{Feature} 
	& \colhead{$\lambda_{\rm rest}$~(\AA)} 
	& \colhead{$\lambda_{\rm obs}$~(\AA)} 
	& \colhead{$F_{\rm obs}$\tablenotemark{\dag}} 
	& \colhead{FWHM$_{\rm obs}$~(\kms)} 
	& \colhead{$W_\lambda^{\rm rest}$~(\AA)}
    }
\startdata
% Spectral Feature       lam_rest  lam_obs          flux       FWHM(obs) E.W.rest
%
H$\alpha^\ddag$          & 6562.79 & 16417.08 & 92.2       & 4112  & 441 \nl
H$\alpha$--narrow$^\ddag$&         & 16414.08       & 26.5       & 2815  &  127 \nl
H$\alpha$--broad$^\ddag$ &         & 16420.07       & 65.8       & 6851  &  314 \nl
H$\beta^\ddag$           & 4861.33 & 12154.17       & 16.0       & 5183  &   63 \nl
H$\gamma$               & 4340.47 & --             & 0$\pm$0.11 & --    & --   \nl
H$\delta$               & 4101.74 & --             & 0$\pm$0.17 & --    & --   \nl
[\ion{O}{3}]            & 5007    & --             & 0$\pm$0.03 & --    & --   \nl
[\ion{O}{2}]            & 3727    & --             & 0$\pm$0.17 & --    & --   \nl
\FeII\ ($J$-band)       & 4400-5400    &           & 23.0       &       &      \nl
\FeII\ ($H$-band)       & 5800-7000    &           & 2.43       &       &      \nl
\enddata
\label{emdata}
\tablenotetext{\dag}{All fluxes are in units of $10^{-15}\ {\rm
erg\ s^{-1}\ cm^{-2}}$.}
\tablenotetext{\ddag}{Properties measured from \FeII-subtracted spectrum.}
\end{deluxetable}

%%%%%%%%%%%%%%%%%%%%%%%%%%%%%%%%%%%%%%%%%%%%%%%%%
%% FIGURE CAPTIONS                    %%%%%%%%%%%
%%%%%%%%%%%%%%%%%%%%%%%%%%%%%%%%%%%%%%%%%%%%%%%%%

%\centerline{\bf Figure Captions}
%\medskip

\end{document}